\documentclass[10pt,conference]{IEEEtran}
\IEEEoverridecommandlockouts
\usepackage{caption}
\usepackage{cite}
\usepackage{amsmath,amssymb,amsfonts}
\usepackage{algorithmic}
\usepackage{graphicx}
\usepackage{setspace}
\usepackage{textcomp}
\usepackage{xcolor}
\usepackage{algorithm,algorithmic}
\def\BibTeX{{\rm B\kern-.05em{\sc i\kern-.025em b}\kern-.08em
    T\kern-.1667em\lower.7ex\hbox{E}\kern-.125emX}}
\usepackage{tabularx}
\DeclareUnicodeCharacter{2212}{\textendash}
\makeatletter
\g@addto@macro\normalsize{%
 \setlength\abovedisplayskip{2pt}
 \setlength\belowdisplayskip{2pt}
 \setlength\abovedisplayshortskip{2pt}
 \setlength\belowdisplayshortskip{2pt}
}
\usepackage{titlesec}
 \usepackage{url}
\titlespacing{\section}{0pt}{2pt}{0pt}
\allowdisplaybreaks
\usepackage{bbm}
\usepackage[mathscr]{euscript}
 \let\mathscr\relax
\usepackage[scr]{rsfso}

\newcommand{\sfur}{\mathsf{ur}}
\newcommand{\sfem}{\mathsf{em}}
\newcommand{\pro}{\mathsf{pro}}
\newcommand{\tx}{\mathsf{tx}}

\newcommand{\st}{\mathsf{s.t.}}
\DeclareMathOperator*{\argmax}{arg\,max}

\begin{document}

\title{On Deep Reinforcement Learning for Traffic Steering Intelligent ORAN}
\author{\IEEEauthorblockN{Fatemeh Kavehmadavani$^1$, Van-Dinh Nguyen$^2$, Thang X. Vu$^1$, and Symeon Chatzinotas$^1$}
\IEEEauthorblockA{$^1$\textit{Interdisciplinary Centre for Security, Reliability
and Trust (SnT), University of Luxembourg} \\
$^2$\textit{College of Engineering \& Computer Science, VinUniversity, Vietnam} \\
 Email: \{fatemeh.kavehmadavani, thang.vu, symeon.chatzinotas\}@uni.lu; dinh.nv2@vinuni.edu.vn}
\thanks{Partially supported by the ERC AGNOSTIC project (ref. EC/H2020/ERCAdG/742648/AGNOSTIC), and the Luxembourg National Fund via project FNR ASWELL, ref. FNR/C19/IS/13718904/ASWELL.}
\vspace{-25pt}}

\maketitle

\begin{abstract}
This paper aims to develop the intelligent traffic steering (TS) framework, which has recently been considered as one of the key developments of 3GPP for advanced 5G. Since achieving key performance indicators (KPIs) for heterogeneous services may not be possible in the monolithic architecture, a novel deep reinforcement learning (DRL)-based TS algorithm is proposed at the non-real-time (non-RT) RAN intelligent controller (RIC) within the open radio access network (ORAN) architecture. To enable ORAN's intelligence, we distribute traffic load onto appropriate paths, which helps efficiently allocate resources to end users in a downlink multi-service scenario. Our proposed approach employs a three-step hierarchical process that involves heuristics, machine learning, and convex optimization to steer traffic flows. Through system-level simulations, we show the superior performance of the proposed intelligent TS scheme, surpassing established benchmark systems by $45.50\%$.
\end{abstract}

\section{Introduction}\label{section1}
The emergence of fifth-generation (5G) cellular networks has introduced new service classes, \textit{namely} ultra-reliable low-latency (uRLLC) and enhanced mobile broadband (eMBB) services \cite{saad2019vision}. The current 5G architecture is inadequate to support diverse and competing services with limited resources. To overcome this challenge, a transition to a disaggregated architecture is necessary for the advancement of 5G and future sixth-generation (6G) networks. The open radio access network (ORAN) has emerged as a promising solution, emphasizing \textit{intelligence} and \textit{openness} \cite{niknam2020intelligent}.

ORAN employs functional splitting, dividing the base station functions into radio unit (RU), distributed unit (DU), and central unit (CU) according to 3GPP standards. It also integrates the near-real-time (near-RT) RAN intelligent controller (RIC) and non-real-time (non-RT) RIC modules at the management and control layers, introducing intelligence and closed control loops for autonomous actions and periodic feedback. This enables RAN optimization and the implementation of machine learning/artificial intelligence  (ML/AI) solutions, creating adaptive and intelligent radio access network (RAN) layers within the ORAN framework. In 5G wireless networks, the traffic steering (TS) scheme, as the first user-specific ORAN intelligent handover framework, plays a crucial role in connecting heterogeneous network frameworks to multiple radio access technologies (RATs) and RAN components. It allows intelligent handover decisions based on feedback-driven analysis of network states and performance across various ORAN's components. When considering traffic preferences, the TS scheme offers great potential to improve overall network performance.

Efficient data flow management is crucial in 5G networks to meet diverse service requirements. To this end, this study utilizes network slicing (NS) and multi-connectivity (MC) technologies to improve data rates for eMBB services and reduce latency for uRLLC services \cite{beschastnyi2020modelling}. This research explores the integration of mixed numerologies in the frequency domain, benefiting the mini-slots concept to support latency-critical applications (\textit{i.e.}, uRLLC). This enhances the flexibility of RAN slicing, enabling efficient and dynamic resource management.

Despite extensive research on TS in 4G and LTE-advanced networks, there is a lack of literature that specifically addresses TS in 5G networks. The authors in \cite{burgueno2020traffic} proposed the TS-based MC scheme to improve the quality of experience of the eMBB services while reducing network expenses. The reinforcement learning (RL) was studied in \cite{priscoli2020traffic} model network selection and TS in 5G networks, focusing on load balancing and QoS requirements. Additionally, the work \cite{dryjanski2016unified} investigated a unified TS scheme to optimize resource utilization. However, there has been limited research on TS modeling specifically within the ORAN architecture. In our previous work \cite{kavehmadavani2023intelligent}, we investigated a slice isolation mechanism for allocating RAN resources in the ORAN architecture to handle non-uniform traffic steering. 

Existing research has overlooked the intricate challenges associated with decision-making per time slot in the presence of unknown channel state information (CSI). To bridge this gap, we present a comprehensive TS framework that leverages deep reinforcement learning (DRL). This framework empowers automated networks to reduce computational complexity by making decisions per frame instead of every time slot, while addressing incomplete knowledge of CSI. DRL is developed as an intelligent agent to efficiently manage traffic steering while accommodating constraints of limited initial information and the inherent computational complexity in the binary allocation problem.

In this study, our goal is to develop a DRL-based TS scheme that incorporates slice-aware RAN slicing, dynamic MC technique, and mixed numerologies, aiming to achieve the optimal steering of traffic flows. In summary, our key contributions are outlined as follows:
\begin{itemize}
    \item We formulate a joint optimization problem of flow-split distribution, congestion control, and scheduling scheme befitting the ORAN architecture. The proposed problem takes into account dynamic MC, slice-aware RAN slicing, and mixed-numerologies in the frequency domain, subject to QoS requirements of both eMBB and uRLLC traffics. 
    \item To account for the mini-slot concept, the proposed problem is executed on two different time scales (frame and mini-slot). This division results in two subproblems (long-term and short-term), which are solved at non-RT RIC and near-RT RIC, respectively. This paper introduces a new approach to address the challenge of incomplete information such as CSI and computational complexity. The proposed solution involves the implementation of a double deep Q-network (DDQN) model within the non-RT RIC. This model aims to predict resource block (RB) assignments for each frame instead of TTI, thus improving efficiency and reducing complexity.
    \item Numerical results are presented and compared with benchmark schemes. The effectiveness of our approach is demonstrated through a notable performance improvement of $45.50\%$ in terms of throughput.
\end{itemize}

\section{System Model}\label{section2}
\begin{figure}[t]
  \centering
  \includegraphics[width=0.5\textwidth,trim=4 4 4 4,clip=true]{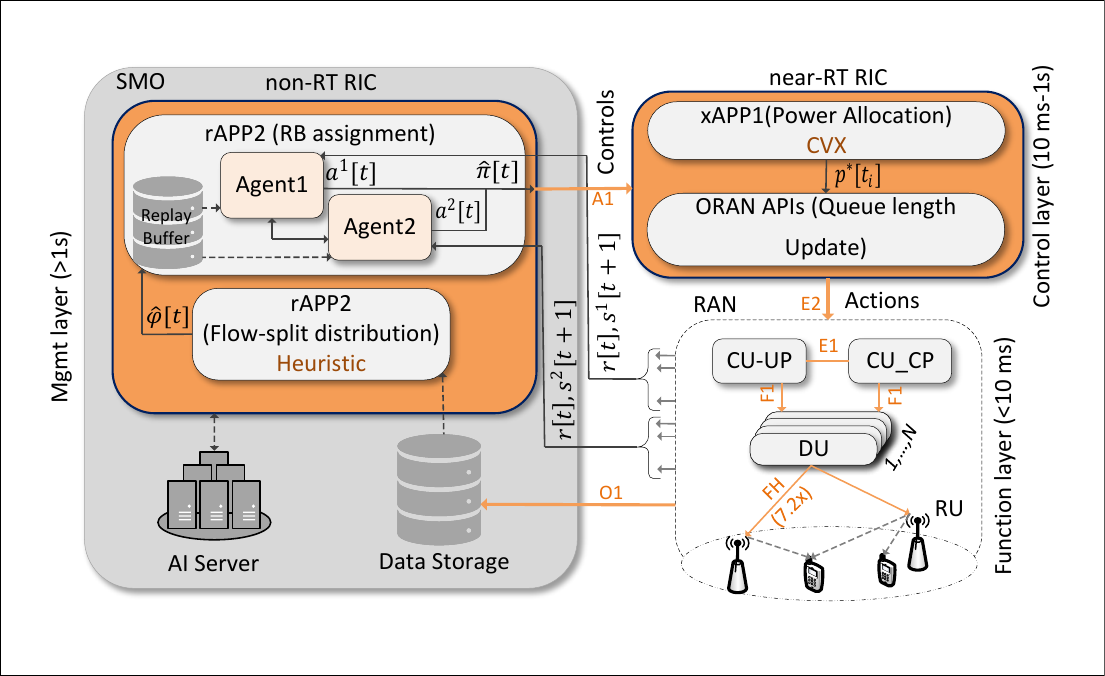} 
  \captionsetup{skip=1pt} 
  \caption{ORAN architecture and ML application workflow}
  \label{fig1}
\end{figure}

As illustrated in Fig. \ref{fig1}, a downlink orthogonal frequency division multiple access (OFDMA) system is considered in the RAN layer, including a set of $M$ multi-antenna RUs denoted $\mathscr{M}\triangleq\{1,\dots, m, \dots, M\}$. Each RU serves a set of $U$ single-antenna users denoted as $\mathscr{U}\triangleq\{1,\dots, u, \dots,U\}$. Users are divided into two non-overlapping sets of $U^{\sfem}$ eMBB and $U^{\sfur}$ uRLLC, \textit{i.e.}, $\mathscr{U}\triangleq \mathscr{U}^{\sfem} \cup \mathscr{U}^{\sfur}$. This paper utilizes RAN resource slicing and MC technology to achieve strict uRLLC latency and high eMBB data rate, owing to different packet sizes (\textit{i.e.}, large eMBB packet size $Z^{\sfem}$ and small uRLLC packet size $Z^{\sfur}$).
This study incorporates mixed-numerology multiplexing in the frequency domain and utilizes a mini-slot-based framework, allowing each RU to allocate time-frequency radio RBs to serve multiple users, thereby enhancing system flexibility \cite{3gpp2017}. In the discrete-time system, each frame is denoted by $t \in \{1, 2, \dots, T\}$. Within each frame, there are $T_i=\Delta/\delta_i$ transmission time intervals (TTIs) indexed as $t_i$ in the $i$-th numerology, where $\Delta$ and $\delta_i$ denote the frame duration, and the duration of each TTI, respectively. 

We consider two numerologies per slice, indexed as $i=1$ and $i={2}$ for the eMBB and uRLLC services, respectively \cite{kihero2019inter}. The system bandwidth (BW) $B$ is divided into two independent BW parts by the split variable $\alpha\in(0,1)$. This makes two slices with BW of $B_i|_{i=1} = (1-\alpha)B$ and $B_i|_{i=2} = \alpha B - B_{G}$ per slice including $F_i= \lfloor{B_i}/{\beta_i} \rfloor$ RBs, where the guard band $B_{G}=180$ kHz helps reduce inter-numerology interference within adjacent sub-bands. Let $\beta_i$ be each RB's BW.  

To handle users' packets, we employ the $M/M/1$ processing queue model for service. Due to the MC configuration, the $u$-th data flow is split into sub-flows by CU. These sub-flows can be transmitted through a maximum of $M$ paths and then aggregated at the intended user. The global flow-split decision, denoted by $\boldsymbol {\varphi} [t] \triangleq \{\boldsymbol{\varphi}_u[t]; \forall u | \sum_{m}\varphi_{m,u}[t]=1, \varphi_{m,u}[t] \in [0,1]\}$, determines the portion of the data flow routed to the user $u$ via RU $m$ in time-frame $t$. The flow-split portion vector of user $u$ is represented by $\boldsymbol{\varphi}_u [t] \triangleq \Big [\varphi_{m,u}[t] \Big]^T$, with $\sum_{m}\varphi_{m,u}[t]=1$ and $\varphi_{m,u}[t] \in [0,1]$ indicating the proportion of data flow transmitted via RU $m$ to user $u$ per time-frame $t$.

By applying the Shannon-Hartley theorem, the downlink data rate of the $u$-the eMBB user served by RU $m$ at $t_i$ can be modelled as 
\begin{IEEEeqnarray}{lCr}\label{eq1}
    R_{m,u}^{\sfem}(\boldsymbol{p}^{\sfem}[t_i])= \sum_{f_i,i}\beta_i \log_{2}\Bigl(1+\frac{p_{m,u,f_i}^{\sfem}[t_i] g_{m,u,f_i}[t_i]}{N_{0}}\Bigl)\quad
\end{IEEEeqnarray}
where $N_0$ and $g_{m,u,f_i}[t_i] \triangleq \|\boldsymbol{h}_{m,u,f_i}[t_i]\|^2_2$ are the AWGN's power and the effective correlated channel gain, respectively;  $p_{m,u,f_i}^{\sfem}[t_i]$ denotes the transmit power from RU $m$ to eMBB user $u$ at sub-band $f_i$ and  TTI $t_i$. We denote by $\boldsymbol{G}[t] \triangleq [g_{m,u,f_i}[t_i]]^T$ the channel gain between all RUs' RB($t_i, f_i$) to all users in time-frame $t$. Thanks to Big-M formulation theory, we can avoid non-convexity issues in (\ref{eq1}). We consider the scheduling constraint: $0 \leq p_{m,u,f_i}^{\sfem}[t_i] \leq \pi_{m,u,f_i,t_i}^{\sfem}[t]P^{\max}_m$ to ensure that if $\pi_{m,u,f_i,t_i}^{\sfem}[t]=0$ then $p_{m,u,f_i}^{\sfem}[t_i]=0$, where $P^{\max}_m$ is the maximum available transmission power of RU $m$. Besides, constraint $\sum_{t_i, i}R_{m,u}^{\sfem}(\boldsymbol{p}^{\sfem}[t_i])\geq \varphi_{m,u}[t] \lambda^{\sfem}_u[t] Z^{\sfem} \Delta$ is refereed to eMBB QoS requirements, where $\lambda^{\sfem}_u[t]$, and $\varphi_{m,u}[t]\lambda^{\sfem}_u[t]Z^{\sfem}$ [bits/frame] represent the eMBB arrival traffics in time-frame $t$, and the sub-flow of $u$-th eMBB user in RU $m$, respectively. It ensures that the achievable data rate of $u$-th eMBB user from RU $m$ meets the estimated value of $\varphi_{m,u}[t]$.

Note that $\pi_{m,u,f_i,t_i}^{\mathsf{x}}[t] \in \{0,1\}$ denotes the binary variable to indicate whether RB($f_i,t_i$) associated with sub-band $f_i$ in TTI $t_i$ of RU $m$ in time-frame $t$ is allocated to the user $u$-th eMBB/uRLLC service,  satisfying orthogonality constraints, where $\mathsf{x} \in \{\sfem, \sfur\}$. If RB($f_i,t_i$) is assigned to the $u$-th eMBB/uRLLC user via RU $m$, we have $\pi_{m,u,f_i,t_i}^{\mathsf{x}}[t]=1$; otherwise $\pi_{m,u,f_i,t_i}^{\mathsf{x}}[t]=0$. Let define the RB assigned matrix as $\boldsymbol{\pi}^{\mathsf{x}}[t] \triangleq \Big[ \boldsymbol{\pi}^{\mathsf{x}}_{m,u}[t]\Big]^T$, where $\boldsymbol{\pi}^{\mathsf{x}}_{m,u}[t] \triangleq \Big[\pi^{\mathsf{x}}_{m,u,f_i,t_i}[t] \Big]^T$ for the eMBB/uRLLC services.

The maximum achievable rate that the $u$-th uRLLC user may achieve from RU $m$ at a certain block-length and error probability ($P_e$) is roughly represented by
{\small{\begin{IEEEeqnarray}{rCl}\label{eq2}
    \nonumber R_{m,u}^{\sfur}(\boldsymbol{p}^{\sfur}[t_i],\boldsymbol{\pi}^{\sfur}[t])=&\sum_{f_i,i}\beta_i \Big[\log_{2}\Big(1+\frac{ p_{m,u,f_i}^{\sfur}[t_i] g_{m,u,f_i}[t_i] }{N_{0}}\Big)\\
    &-\log_{2}(e)\pi_{m,u,f_i,t_i}^{\sfur}[t]\Psi\Big]
\end{IEEEeqnarray}}}
where $\Psi\triangleq\frac{\sqrt{V} Q^{-1}(P_{e})}{\sqrt{ \delta_i \beta_i}}$, $V$, and $Q^{-1}$ are the channel dispersion and the inverse of the Gaussian Q-function, respectively. Based on the Big-M formulation theory, we approximate $V\approx1$ under the constraint uRLLC SNR $\Gamma_0 \geq 5$dB. In other words, it follows that $\frac{N_0 \Gamma_0}{g_{m,u,f_i}[t_i]} \pi_{m,u,f_i,t_i}^{\sfur}[t] \leq p_{m,u,f_i}^{\sfur}[t_i] \leq \pi_{m,u,f_i,t_i}^{\sfur}[t] P_m^{\max}$  \cite{schiessl2015delay}. We define the power allocation vector of eMBB/uRLLC traffic as $\boldsymbol{p}^{\mathsf{x}}[t_i]\triangleq[p_{m,u,f_i}^{\mathsf{x}}[t_i]]^T$. Similarly to the eMBB service,  the achievable data rate of the $u$-th uRLLC user from RU $m$ meets the estimated value of $\varphi_{m,u}[t]$ as $\sum_{t_i, i}R_{m,u}^{\sfur}(\boldsymbol{p}^{\sfur}[t_i],\boldsymbol{\pi}^{\sfur}[t])\geq \varphi_{m,u}[t] \lambda^{\sfur}_u[t] Z^{\sfur} \Delta$, where $\lambda^{\sfur}_u[t]$ is the arrival traffics of the $u$-th uRLLC user in time-frame $t$. It implies that every RB allocated to the $u$-th uRLLC user must transmit a complete data packet of size $Z^{\sfur}$.

The queue length of the $u$-th data flow in RU $m$ is defined as $q_{m,u}[t_i]= \big(q_{m,u}[t_i-1]+\varphi_{m,u}[t]\lambda^{\mathsf{x}}_u[t]Z^{\mathsf{x}} \Delta- R_{m,u}^{\mathsf{x}}(\boldsymbol{p}[t_i])\delta_i\big)^+$ [bits]. Where, $(x)^+ \triangleq \max\{x,0\}$. To maintain a maximum buffer size of $q^{\max}$, we impose the constraint $ \boldsymbol{q}_{m}[t] \leq q^{\max}$, where $\boldsymbol{q}_{m}[t] = \sum_{t_i=1}^{T_i} \sum_{u} q_{m,u}[t_i]$.

The uRLLC end-to-end (e2e) latency of the $u$-th uRLLC user at time-frame $t$ can be expressed as \cite{kavehmadavani2023intelligent}
\begin{IEEEeqnarray}{rCl}\label{eq3} 
\nonumber \tau_{u}^{\sfur}[t]=&&\;\tau^{\pro}_{cu}[t]+\tau^{\tx}_{cu,du}[t]+\tau^{\pro}_{du}[t]\\
&&+\; \tau_{du,ru}^{\tx}[t] +\tau_{ru,u}^{\tx}[t]+ \tau^{\pro}_{ru}[t]\qquad
\end{IEEEeqnarray}
where $\tau^{\pro}_{cu}[t]$, $\tau^{\pro}_{du}[t]$ and $\tau^{\pro}_{ru}[t]$ are the CU, DU and RU processing times, respectively; $\tau^{\tx}_{cu,du}[t]$, $\tau^{\tx}_{du,ru}[t]$ and $\tau^{\tx}_{ru,u}[t]$ are the  transmission latency under the midhaul (MH), fronthaul (FH) and RU-user communication latency at time-frame $t$, respectively. Thanks to the RAN slicing concept, the proposed system consistently possesses the necessary resources to instantly serve the uRLLC upon arrival, thereby ensuring that the uRLLC experiences fewer delays in queueing. Thus, the transmission time of the RU-user links becomes the main factor against reaching the tight uRLLC latency requirement. The latency  $\tau_{ru,u}^{\tx}[t] = \delta_i \times \argmax_{t_i} \{\pi^{\sfur}_{m,u,f_i,t_i}[t]\}$  is calculated as the time difference (measured in TTI) between the moment a uRLLC packet enters the buffer and the moment it is scheduled and transmitted from the buffer. To satisfy the minimum latency requirement for the $u$-th uRLLC user, the end-to-end latency is constrained by a predefined threshold of $D^{\sfur}$, such that $\tau_u^{\sfur}[t] \approx \tau_{ru,u}^{\tx}[t] \leq D^{\sfur}$.

\section{Deep Reinforcement Learning-aided Intelligent Traffic Steering}\label{section3}

\subsection{Problem Formulation}
\vspace{-5pt}
\noindent \textbf{Utility function:} We aim to optimize the performance of the ORAN by jointly considering flow split distribution, congestion control, and scheduling for eMBB and uRLLC services subject to QoS requirements, power budget, slice awareness, and other practical constraints. The utility function is designed to simultaneously address eMBB throughput and worst-user e2e uRLLC latency, \textit{i.e.}, \begin{align}
\omega \sum_{u \in \mathscr{U}^{\sfem}}\frac{\Bar{q}^{\sfem}_{u}}{q_0} + (1-\omega)\max_{u\in\mathscr{U}^{\sfur}}\frac{\Bar{\tau}^{\sfur}_{u}}{\tau_0}
\end{align}
where $\Bar{q}^{\sfem}_u \triangleq \lim_{t_i \rightarrow \infty} \frac{1}{t_i} \sum_{\tau =1}^{t_i} \sum_{m} q_{m,u}[\tau]$ and $\Bar{\tau}^{\sfur}_{u} = \delta_i.\mathbb{E}_t\{\argmax_{t_i}\pi^{\sfur}_{m,u,f_i,t_i}[t]\}$ are the long-term average queue length of $u$-th eMBB data flow and uRLLC latency of $u$-th uRLLC data flow, respectively. Here, we introduce the reference throughput $q_0>0$ and the reference latency $\tau_0>0$ for eMBB and uRLLC, respectively, to balance two objective functions. The priority parameter $\omega\in[0, 1]$ allows prioritization between eMBB and uRLLC. Overall, the intelligent TS optimization problem is mathematically formulated as 
\begin{subequations} \label{op1}
\begin{IEEEeqnarray}{cl}
     &\min_{\boldsymbol{\varphi}, \boldsymbol{\pi},\boldsymbol{p}} \omega \sum_{u \in \mathscr{U}^{\sfem}}\frac{\Bar{q}^{\sfem}_{u}}{q_0}+ (1-\omega)\max_{u\in\mathscr{U}^{\sfur}}\frac{\Bar{\tau}^{\sfur}_{u}}{\tau_0}\\
    &\st \quad \pi_{m,u,f_i,t_i}^{\mathsf{x}}[t]\in \{0,1\}; \ \forall t, \mathsf{x} \in \{\sfem, \sfur\}\label{op1:b}\\
    & \qquad \sum_{m,u}\big(\pi_{m,u,f_i,t_i}^{\sfem}[t]+\pi_{m,u,f_i,t_i}^{\sfur}[t]\big)\leq 1;\ \forall f_i,t_i\label{op1:c}\\
    & \qquad \sum_{t_i = 1}^{D^{\sfur}/\delta_i}\sum_{f_i = 1}^{F_i}\pi_{m,u,f_i,t_i}^{\sfur}[t] \geq e^{\sfur}_{u}[t];\ \forall u\in \mathscr{U}^{\sfur}, i=1 \label{op1:d}\\
    & \qquad \sum_{t_i=1}^{T_i} \sum_{f_i =1}^{F_i}\pi_{m,u,f_i,t_i}^{\sfem}[t] \geq e^{\sfem}_{u}[t]; \ \forall u\in \mathscr{U}^{\sfem}, i=2 \label{op1:e}\\
    & \qquad 0 \leq p_{m,u,f_i}^{\sfem}[t_i]\leq \pi_{m,u,f_i,t_i}^{\sfem}[t] P_m^{\max}; \ \forall t_i \label{op1:f}\\
    &  \nonumber\qquad \frac{N_0 \Gamma_0}{g_{m,u,f_i}[t_i]} \pi_{m,u,f_i,t_i}^{\sfur}[t] \leq p_{m,u,f_i}^{\sfur}[t_i] \leq \pi_{m,u,f_i,t_i}^{\sfur}[t] P_m^{\max} \\
    & \label{op1:g}\\
    & \qquad \sum_{f_i,u,i}^{}(p_{m,u,f_i}^{\sfem}[t_i]+p_{m,u,f_i}^{\sfur}[t_i])\leq P_{m}^{\max}; \ \forall t_i, m\in \mathscr{M} \label{op1:h}\\
    & \qquad \boldsymbol{\varphi}_u [t] \in \boldsymbol{\varphi}[t];\   \forall t, u\in \mathscr{U} \label{op1:i}\\ 
    & \qquad R^{\mathsf{x}}_{m,u}(\boldsymbol{p}^{\mathsf{x}}[t_i])\geq \varphi_{m,u}[t]\lambda^{\mathsf{x}}_{u}[t] Z^{\mathsf{x}} \Delta; \ \forall \mathsf{x} \in \{\sfem, \sfur\} \label{op1:j}\\
    & \qquad \tau^{\sfur}_{u} (\boldsymbol{\pi}^{\sfur}[t])\leq D^{\sfur}; \ \forall u\in \mathscr{U}^{\sfur}\label{op1:l}\\ 
    & \qquad \sum_{t_i}\sum_{u} q_{m,u}[t_i] \leq q^{\max}; \ \forall m\in \mathscr{M} \label{op1:m}
\end{IEEEeqnarray}
\end{subequations}
where $\boldsymbol{\pi}^{\mathsf{x}}[t], \boldsymbol{\varphi}[t]$ and $\boldsymbol{p}^{\mathsf{x}}[t_i]$ are the vectors encompassing the sub-band assignments, flow-split portions, and power allocation vectors at frame $t$ and TTI $t_i$, respectively. Here, the constraint (\ref{op1:c}) is the orthogonality constraint to ensure that each RB of RU is allocated to only one user. To further exploit the existing  slices' RBs, the slice-aware constraints (\ref{op1:d}) and (\ref{op1:e}) are proposed to improve the utilization of radio resources, where $e^{\sfur}_{u}[t]=\lceil (\lambda^{\sfur}_u[t]-\Omega_u[t])^+/2\rceil|_{\forall u \in \mathscr{U}^{\sfur}}$ and $e^{\sfem}_u[t]=(\lfloor((F_i\times T_i)-\sum_{u^{\sfur}} \min(\lambda^{\sfur}_u[t],\Omega_u[t]))/U^{\sfem}\rfloor)^+|_{\forall u \in \mathscr{U}^{\sfem}, i=2}$, in which $\Omega_u[t]= \frac{\lambda^{\sfur}_u[t]}{\sum_{\sfur}\lambda^{\sfur}_u[t]}.\Omega$ is the maximum number of RBs for each uRLLC user in the proprietary slice of uRLLC per time-frame $t$. Here $\Omega = (F_i \times D^{\sfur}/\delta_i)|_{i=2}$ represents the number of RBs available from the dedicated uRLLC slice that meet the uRLLC latency constraint. Finally, the constraint (\ref{op1:h}) ensures that the total transmission power is not greater than the RU power budget $P_m^{\max}$.

\noindent \textbf{Challenges of Solving Problem (\ref{op1}):} Problem \eqref{op1} presents several challenges to be optimally solved due to non-convexity in constraints (\ref{op1:j}) and (\ref{op1:m}) with respect to $\boldsymbol{\varphi}[t]$ and $\boldsymbol{p}^{\mathsf{x}}[t_i]$, as well as the binary nature of $\boldsymbol{\pi}^{\mathsf{x}}[t]$. These characteristics make problem \eqref{op1} a mixed-integer non-linear program (MINLP) and computationally expensive. On the other hand, wireless systems often face dynamic changes in network conditions, such as time-varying channels and fluctuating traffic demands. As a result, standard optimization methods are not applicable to solve the problem directly and efficiently.

Toward an efficient and stable solution, we divide problem (\ref{op1}) into two subproblems on the long-term $t$ scale and short-term time scale $t_i$, respectively. The flow-split decisions and RB assignments are heavily influenced by the RAN layer's reliance on the updated previous states due to the queue length and incomplete knowledge of channels at the start of each frame. The flow-split vector $\boldsymbol{\varphi}[t]$ and the RB assignment vector $\boldsymbol{\pi}^{\mathsf{x}}[t]$ are updated per frame $t$, while the power allocation vector $\boldsymbol{p}^{\mathsf{x}}[t_i]$ is optimized based on the effective real-time CSI in time slot $t_i$, enabling adaptability in dynamic environments.

\subsection{Proposed Three-Steps Methodology for Solving (\ref{op1})} 
As mentioned previously, optimizing long-term variables ($\boldsymbol{\varphi}[t]$ and $\boldsymbol{\pi}[t]$) is challenging due to the unknown queue length and channels at the beginning of the time-frame. Under incomplete information, we determine $\boldsymbol{\varphi}[t]$ and $\boldsymbol{\pi}[t]$ based on observable data. Algorithm \ref{alg1} presents a DRL-based intelligent TS approach to optimize flow split distribution, congestion control, and resource allocation on long- and short-term time scales. 

\begin{algorithm}[!h]
\begin{algorithmic}[1]
\fontsize{9}{9}\selectfont
\protect\caption{Intelligent TS Algorithm for Solving (\ref{op1})}
\label{alg1}
\global\long\def\algorithmicrequire{\textbf{Initialization:}}
\REQUIRE  Set $t_i=1$, $t=1$, $\boldsymbol{\varphi}_u[1]= \frac{1}{M}\boldsymbol{1_{M\times 1}}$, and $\boldsymbol{q}[1]= \big[\boldsymbol{q}_m[1]\big]^T$, where $\boldsymbol{q}_m[1]=\boldsymbol{0}; \forall m$. 
\FOR {$t=1,2, \dots,T$} 
\STATE \texttt{\textbf{Traffic flow splitting estimation:}} The embedded heuristic method deployed in rAPP1 splits the traffic flows of all users $\hat{\boldsymbol{\varphi}}[t]$ for time-frame $t$ by (\ref{eq5});
\STATE \texttt{\textbf{RB assignment prediction:}} Given the sorted data ($\boldsymbol{\lambda}[t], \boldsymbol{\hat{\mathrm{\varphi}}}[t], \boldsymbol{q}[t-1], \boldsymbol{G}[t-1], \boldsymbol{e}^{\mathsf{x}}[t]$) in data storage, the rAPP2 consists of two DRL agents predicts the binary RB assignments $\hat{\boldsymbol{\pi}}[t]$ for time-frame $t$ via Algorithm \ref{alg2}, where $\boldsymbol{e}^{\mathsf{x}}[t]=[\boldsymbol{e}_u^{\mathsf{x}}[t]]^T$;
    \FOR {$t_i=1,2, \dots,T_i$}
    \STATE \texttt{\textbf{Optimizing power allocation:}} Given the vector of the queue length $\boldsymbol{q}[t]$, and two predicted long-term variables: ($\hat{\boldsymbol{\varphi}}[t]$, and $\hat{\boldsymbol{\pi}}[t]$), solve the problem (\ref{op2}) to obtain the power allocation $\boldsymbol{p}^*[t_i]$;
    \STATE \texttt{\textbf{Updating queue-lengths:}} Queue-lengths are updated as:
    $q_{m,u}[t_i] = \big(\big[q_{m,u}[t_i-1] + \hat{\varphi}_{m,u}[t] \lambda^{\mathsf{x}}_u[t] Z^{\mathsf{x}} \delta_i - R_{m,u}^{\mathsf{x}} [\boldsymbol{p}^*[t_i]]\delta_i\big]\big)^+$, where $\mathsf{x} \in \{\sfur, \sfem\}$
    \ENDFOR
\STATE Update $\{\boldsymbol{\lambda}[t], \boldsymbol{\hat{\mathrm{\varphi}}}[t], \boldsymbol{q}[t\text{-}1], \boldsymbol{G}[t \text{-} 1], \boldsymbol{e}^{\mathsf{x}}[t]\}:= \{\boldsymbol{\lambda}[t+1], \boldsymbol{\hat{\mathrm{\varphi}}}[t+1], \boldsymbol{q}[t], \boldsymbol{G}[t], \boldsymbol{e}^{\mathsf{x}}[t+1]\}$;
\ENDFOR 
\end{algorithmic}
\end{algorithm}

\noindent \textbf{Heuristic method:} The non-RT RIC-based rAPP1 employs a heuristic-based approach to estimate the flow-split decision $\boldsymbol{\varphi}[t]$ for traffic flow separation. To handle the unpredictable data arrival rate in future frames, we utilize a moving average of observed rates from recent TTIs. Let $\Bar{R}^{\mathsf{x}}_{m,u}[t]= \frac{1}{W} \sum_{l=t-W+1}^{t} R^{\mathsf{x}}_{m,u}[l]$ be the achievable rate of the $u$-th generic user served by RU $m$ in time-frame $l$, and $W$ is the window size, where $R^{\mathsf{x}}_{m,u}[l]$. The data flow of the $u$-th user to the $m$-th RU can be split as
\begin{equation}\label{eq5}
\hat{\varphi}_{m,u}[t] = \frac{\Bar{R}^{\mathsf{x}}_{m,u}[t]}{\sum_{m} \Bar{R}^{\mathsf{x}}_{m,u}[t]},\ \forall m, u, \mathsf{x} \in \{\sfem,\sfur\}
\end{equation}
The estimated flow-split decision $\hat{\boldsymbol{\varphi}}[t] = \big[\hat{\varphi}_{m,u}[t]\big]^T$ is promptly transferred to rAPP2 embedded at non-RT RIC to predict RB assignment $\boldsymbol{\pi}^{\mathsf{x}}[t]$.

\noindent \textbf{Double Deep Q-Network (DDQN):} Unlike previous works that assign RBs per each TTI, this study employs a DRL-based approach to predict the RB assignment $\boldsymbol{\pi}^{\mathsf{x}}[t]$ at the beginning of each frame. This helps enable the dynamic scheduling of multi-services and reduces computational complexity. To address such a complex optimization problem, this paper utilizes a cooperative multi-agent system where each agent per slice interacts with the environment. Each agent receives a subset of the environment observations and takes a subset of actions, resulting in more accurate decision-making and improved network performance in dynamic multi-action environments. Since deep Q-Networks (DQNs) are designed to large-scale state space and fitted to discrete action space, this model is selected to predict the RB assignment vector in such proposed networks. To overcome overestimation and slow convergence issues in these models, this study adopts the double DQN (DDQN) approach for each agent, which could be generalized to multi-action binary scenarios by modifying the architecture and output layer of the neural network accordingly. By separating the max operation in the target network into action selection and evaluation, DDQN reduces overestimations and improves value estimation.

Each agent including the DDQN model decouples action selection from action evaluation by defining two \textit{evaluation} and \textit{target} neural networks. While action selection and policy evaluation are performed in the evaluation network $Q(\boldsymbol{\mathrm{s}},\boldsymbol{\mathrm{a}};\boldsymbol{\theta}^{Q})$, the target network $Q(\boldsymbol{\mathrm{s}},\boldsymbol{\mathrm{a}};\boldsymbol{\theta}^{\mu})$ calculates the future Q value. Note that $\boldsymbol{\theta}^{Q}$ and $\boldsymbol{\theta}^{\mu}$ show the trainable parameters (weights and biases) of the evaluation and target neural networks, respectively. The target network is updated every $C$ steps and optimizes $\boldsymbol{\theta}$ by minimizing the mean square loss: $\mathcal{L}(\boldsymbol{\theta})= \mathbb{E}\big(y - Q(\boldsymbol{\mathrm{s}},\boldsymbol{\mathrm{a}};\boldsymbol{\theta}^{Q})\big)^2$, where $y= \boldsymbol{\mathrm{r}}+\gamma Q(\boldsymbol{\mathrm{s}}',\argmax_{\boldsymbol{\mathrm{a}}}Q(\boldsymbol{\mathrm{s}}',\boldsymbol{\mathrm{a}};\boldsymbol{\theta}^{Q});\boldsymbol{\theta}^{\mu})$, with $\boldsymbol{\mathrm{r}}$ and $\boldsymbol{\mathrm{s}}'$ being the reward and the new state, respectively. Besides, DDQN uses the concept of training neural networks using random batches stored in \textit{replay memory} to stabilize the learning model and remove correlations between observations. Hence, the transition $(\boldsymbol{\mathrm{s}}, \boldsymbol{\mathrm{a}}, \boldsymbol{\mathrm{r}}, \boldsymbol{\mathrm{s}}')$ per each slice is stored in the replay memory data set $D$ based on the first-come-first-serve buffer with limited capacity to be used in the training phase. The summary of the proposed learning method is given in Algorithm \ref{alg2}.

\noindent\textit{State, Action Spaces and Reward Function}: In our cooperative multi-agent system, each agent operates within its own state and action space. The state space encompasses the specific subset of environment observations (slice) accessible to each agent, while the action space comprises the individual set of actions available to each agent. However, it is important to note that the combined action of both agents affects the system's overall dynamics. By defining separate state and action spaces, agents can tailor their perception and interaction with the environment while simultaneously collaborating towards a shared objective.

The state vectors $\boldsymbol{\mathrm{s}}^i[t]|_{i=1,2}$ in the time-frame $t$ are composed of the traffic demand vector $\boldsymbol{\lambda}[t]$, the estimated flow-split distribution $\boldsymbol{\varphi}[t]$, the previous queue length vector $\boldsymbol{q}$[$t$-$1$], the channel gain matrix of each slice $\boldsymbol{G}^i$[$t$-$1$]$|_{i=1,2}$, and $\boldsymbol{e}^{\mathsf{x}}[t]$. Let us define \textit{\textit{i.e.}}, $\mathcal{S}:= \{\boldsymbol{\mathrm{s}}^i[t]|_{i=1,2}| \boldsymbol{\mathrm{s}}^1[t] = (\boldsymbol{\lambda}[t], \boldsymbol{\mathrm{\varphi}}[t], \boldsymbol{q}[t\text{-}1], \boldsymbol{G}^1[t\text{-}1], \boldsymbol{e}^{\sfur}[t]), \boldsymbol{\mathrm{s}}^2[t] = (\boldsymbol{\lambda}[t], \boldsymbol{\mathrm{\varphi}}[t], \boldsymbol{q}[t\text{-}1], \boldsymbol{G}^2[t\text{-}1], \boldsymbol{e}^{\sfem}[t])$. The overall action space is defined as $\mathcal{A} := \{\boldsymbol{\mathrm{a}}^i[t]|_{i=1,2}| \boldsymbol{\mathrm{a}}^1[t] = \big[\pi_{m,u,f_i,t_i}^{\mathsf{x}}[t]\big]^T|_{i=1}, \boldsymbol{\mathrm{a}}^2[t] = \big[\pi_{m,u,f_i,t_i}^{\mathsf{x}}[t]\big]^T|_{i=2}\}$. In this space, $\boldsymbol{\mathrm{a}}[t]$ represents a combination of actions ($\boldsymbol{\mathrm{a}}^i[t]|_{i=1,2}$) taken by each agent. 

To create an effective reward function, this study employs a penalty-based approach that integrates constraints related to the agent's actions ((\ref{op1:c})-(\ref{op1:e}), (\ref{op1:l})). The reward function should suggest a critical evaluation for RB assignment $\boldsymbol{\pi}^{\mathsf{x}}[t]$ in terms of how it will affect the utility of eMBB and uRLLC. Violations of specified constraints incur penalties ($\mathtt{negative\, values}$) to discourage undesirable behavior, while satisfying all constraints rewards the agent with positive reinforcement. This incentivizes decision-making aligned with established constraints and promotes the achievement of system objectives. According to the previously mentioned queue length equation, minimizing the eMBB queue length in the utility function is equivalent to maximizing the eMBB data rate. As a result, we define the reward $\boldsymbol{\mathrm{r}}[t]$ as $\omega\big(\frac{\sum_{t_i,m,u\in \mathscr{U}^{\sfem}} \mathcal{R}^{\sfem}_{m,u}(\boldsymbol{p}^{\sfem}[t_i])}{R_0}\big)-(1-\omega)\big(\frac{\max_{u\in \mathscr{U}^{\sfur}}\{\tau_{u}^{\sfur}[t]\}}{\tau_0}\big)$.
To compute the reward value, it is necessary to solve the short-term power control subproblem.

\noindent \textbf{Short-term Subproblem:} After verifying QoS requirements in Steps 10-19 of Algorithm \ref{alg2}, the subsequent step is to solve the power control problem in the xAPP at near-RT RIC as
\begin{subequations} \label{op2}
\begin{IEEEeqnarray}{cl}
    &\min_{\boldsymbol{p}} \sum_{u \in \mathscr{U}^{\sfem}}\Bar{q}^{\sfem}_{u}\\
    &\st \quad \eqref{op1:f},\eqref{op1:g},\eqref{op1:h},\eqref{op1:j},\eqref{op1:m}. \label{op2:b}
\end{IEEEeqnarray}
\end{subequations}

Since (\ref{op2}) is a convex program, the standard methods can efficiently solve to obtain the optimal transmission power $\boldsymbol{p}^*[t]$.

\section{Numerical Results}
We now numerically evaluate the performance of the proposed algorithms. All users are uniformly located in a circular area with a radius of 500 m. The channels are generated as Rayleigh fading with path loss: $\mathsf{PL}_{\textrm{RU-UE}}=128.1+37.6\log_{10}(d/1000)$. We assume that $u$-th traffic flow of eMBB/uRLLC follows the Poisson distribution with the mean arrival rate of $21.12$ and $1.12$, respectively. Unless otherwise stated, other simulation parameters are given in Table \ref{tab1}. For comparison, we consider the following three benchmark schemes:

\begin{itemize}
\item \textbf{Successive Convex Approximation (SCA)}: Binary variables $\boldsymbol{\pi}^{\mathsf{x}}[t]$ are first relaxed to continuous ones, and then an SCA-based iterative algorithm considering perfect CSI per TTI is developed to solve the approximate convex program \cite{kavehmadavani2023intelligent}. In other words, this scheme serves as the upper bound of the proposed method.
\item \textbf{Fixed-Numerology}: To demonstrate the benefits of flexible numerology in improving system performance, we consider ``Fixed Numerology'' scheme where the subcarrier spacing of 15 kHz is used \cite{korrai2020ran}.
\item \textbf{Uniform-$\boldsymbol{\varphi}$}: This scheme equally splits the data flow, \textit{i.e.}, $\boldsymbol{\varphi}_{u}=\frac{1}{M}; \forall u$, aiming to emphasize the importance of optimizing the flow-split distribution.
\end{itemize}

\begin{algorithm}[]
\begin{algorithmic}[2]
\fontsize{9}{9}\selectfont
\protect\caption{DDQN-based Multi-Agent Algorithm at rAPP2}
\label{alg2}
\global\long\def\algorithmicrequire{\textbf{Initialization:}}
\REQUIRE  Randomly initialize  weights $\boldsymbol{\theta}^{\mu} = \boldsymbol{\theta}^{Q}$, and set replay buffer capacity to $C^{\max}$ and reward value to $\boldsymbol{\mathrm{r}}[t] = 0$.
\FOR {\textit{epoch}} 
\STATE Receive initial observation states for both agents $\boldsymbol{\mathrm{s}}^i[1]|_{i=1,2}$;
    \FOR {$t=1,2, \dots,T$}
    \STATE Generate a random number $\mathrm{rand()}$;
    \IF{$\mathrm{rand()}<\epsilon$}
     \STATE Generate a random action $\boldsymbol{\mathrm{a}}[t]$;
    \ELSE{}
    \STATE Select the action $\boldsymbol{\mathrm{a}}[t]$ that is a joint of  $\boldsymbol{\mathrm{a}}^i[t]|_{i=1,2}$ so that $\boldsymbol{\mathrm{a}}^i[t]|_{i=1,2} = \argmax_{\boldsymbol{\mathrm{a}}^i[t]}Q(\boldsymbol{\mathrm{s}}^i[t],\boldsymbol{\mathrm{a}}^i[t];\boldsymbol{\theta}^{Q})$;
    \ENDIF
    \IF{$\boldsymbol{\mathrm{a}}[t]$ does not satisfy constraints (\ref{op1:d}), (\ref{op1:e}) and (\ref{op1:l})}
    \STATE Set the reward value as $\boldsymbol{\mathrm{r}}[t] += \mathtt{negative\, value}$;
    \ELSE{}
    \STATE Solve problem (\ref{op2}) to get the optimal power allocation $\boldsymbol{p}^*[t]$;
    \IF{It is not feasible}
    \STATE Set the reward value as $\boldsymbol{\mathrm{r}}[t] += \mathtt{negative\, value}$
    \ELSE{}
    \STATE Set the reward value as $\boldsymbol{\mathrm{r}}[t] += \omega \sum_{u,t_i}$ $ \mathcal{R}_u^{\sfem}(\boldsymbol{p}^{\sfem}[t_i])/ R_0-(1-\omega)\max_{u}\{\tau_{u}^{\sfur}[t]\}/ \tau_0$ and observe the new states $\boldsymbol{\mathrm{s}}^i[t+1]|_{i=1,2}$;
    \ENDIF
    \ENDIF
    \STATE Store transition $(\boldsymbol{\mathrm{s}}^i[t],\boldsymbol{\mathrm{a}}^i[t],\boldsymbol{\mathrm{r}}[t],\boldsymbol{\mathrm{s}}^i[t+1])|_{i=1,2}$ in the replay memory $D$;
    \STATE Sample two random mini-batches from replay memory $D$ for training step;
    \STATE Update $\boldsymbol{\theta}^{Q}$ by minimizing the loss function $\mathcal{L}(\boldsymbol{\theta}^Q)$;
    \STATE Update $\boldsymbol{\theta}^{\mu}$ every $C$ steps by resetting $\boldsymbol{\theta}^{\mu}= \boldsymbol{\theta}^{Q}$.
    \ENDFOR
\ENDFOR 
\end{algorithmic}
\end{algorithm}

\begin{figure*}[t]
  \centering
  \includegraphics[width=1\textwidth,trim=2 2 2 2,clip=true]{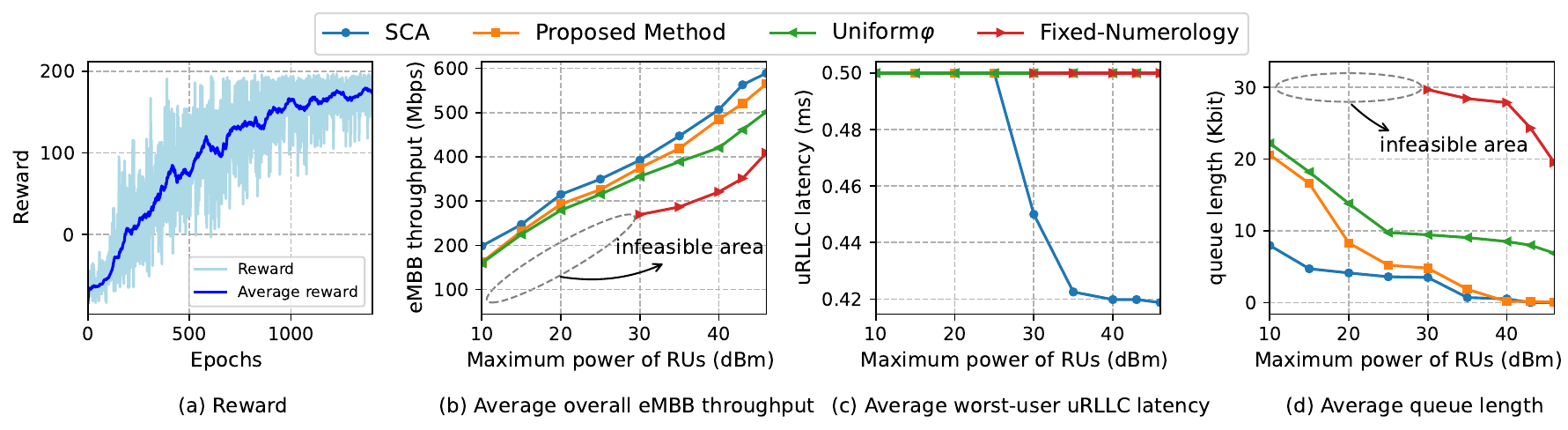} 
  \captionsetup{skip=1pt} 
  \caption{The convergence behaviour of Algorithm \ref{alg2} and performance comparison with existing benchmark schemes}
  \label{fig2}
\end{figure*}

Fig. \ref{fig2}(a) presents a comprehensive performance visualization of Algorithm \ref{alg2} over epochs. It illustrates the agent's rapid adaptation to the dynamic environment, reflecting changes in time-varying channel conditions and arrival packets over time-frames. The same trend can be observed in the average reward, which is steadily increasing during the training episodes. We can also observe that the higher the number of epochs, the higher the average reward that can be obtained.

\begin{table}[!t]
\centering
\caption{Simulation Parameters}
\small
\begin{center}
\scalebox{0.76}{
\begin{tabular}{|l|l|l|l|}
\hline
\textbf{Parameter} & \textbf{Value} &\textbf{Parameter} & \textbf{Value}\\
\hline\hline
 No. of eMBB users & 9 & Required eMBB data rate & 10 Mbps\\
 \hline
 No. of uRLLC users  & 3 & Required uRLLC latency & 0.5 ms\\
 \hline
 No. of RUs  & 4 & Maximum RU's queue-length & 10 KB\\
\hline
 BW of RU & 10 MHz & No. of layers, units & 5, 512 \\
 \hline
 Error probability & $10^{-3}$ & Discount factor & 0.99\\
 \hline
 Power of RU& 46 dBm & Buffer size & 1e+06\\
 \hline
Noise power& -110 dBm & Batch size & 100\\
\hline
uRLLC packet size& 32 B & Soft update coefficient & 0.01\\
\hline
eMBB packet size& 50 KB & Optimizer & \textit{adam}\\
\hline
Length of time-frame& 10 ms & Activation function & \textit{ReLU/softmax}\\
\hline
\end{tabular}
 }
\label{tab1}
\end{center}
\end{table}\raggedbottom

Figs. \ref{fig2}(b)-(d) plot the performance comparison of the proposed scheme with the three benchmark schemes versus the transmit power of the RUs. The results are averaged over 1000 sub-frames. As can be seen, increasing the power budget of the RUs has a positive impact on the sum of eMBB throughput and reduces uRLLC latency and queue length. Fig. \ref{fig2}(b) depicts the system throughput of the eMBB service by varying the maximum RUs' power budgets (from $10$ to $46$ dBm), facilitating the evaluation of different schemes. As expected, the SCA demonstrates superior performance, setting the upper bound for other schemes. The performance gap between Algorithm \ref{alg1} and SCA is less than $2\%$, highlighting the efficiency of DDQN in resource scheduling compared to other benchmark schemes. Compared to the ``Uniform $\boldsymbol{\varphi}$''  and ``Fixed-Numerology'', the proposed scheme offers $10.26\%$ and $45.50\%$ gains at $P^{\max} = 30$ dBm, respectively. Moreover, the fixed-numerology scheme is not feasible at $P^{\max}\leq 30$ dBm. The ``Uniform $\boldsymbol{\varphi}$'' initially performs closely to SCA and our proposed method but declines thereafter. This is attributed to the fact that in the low-power range, UEs with long queues are served by multiple RUs to maximize overall performance. However, beyond this range, a single RU may suffice to serve eMBB traffic.
Fig. \ref{fig2}(c) showcases the worst-user uRLLC latency for different maximum power levels of RUs. As we can see from this figure,  all schemes meet the required uRLLC latency (0.5 ms). It is clear that SCA works better in high power rather than other schemes. The empty region of fixed-numerology at $P^{\max} \leq 30$ dBm shows that the corresponding problem is infeasible. 
Fig. \ref{fig2}(d) depicts the average backlog with different benchmark schemes. As can be seen, the higher the power budget $P^{\max}$, the lower the average queue length. Similar to the previous figures, the results of the proposed method and the SCA are very close to each other, especially at the high power. The fixed-numerology scheme yields the worst performance in terms of the average queue length, whereas the proposed method yields the best one in Fig. \ref{fig2}(d) after SCA. 

\section{Conclusion}
This paper presented an intelligent TS framework to support multi-service scenario. Using multi-connectivity, network slicing, and mixed numerology techniques, the framework efficiently handles distributed traffic load and resource scheduling in a downlink multi-service scenario. To handle dynamic scheduling, time-varying channel states, and reduce computational complexity, we employed DDQN-aided multi-agent model for efficient RB assignment prediction. Extensive simulations show the superior performance of our proposed design over benchmark schemes. Future work entails thoroughly investigating the scalability and adaptability of our intelligent TS framework in alignment with ORAN specifications and 3GPP standardization. 
\begingroup
\setstretch{0.94}
\bibliographystyle{ieeetr}
\bibliography{ref}
\endgroup

\end{document}